\documentclass[numbers,webpdf]{ima-authoring-template}%

\graphicspath{{Fig/}}

\newcommand{\pfr}[2]{\ensuremath{\frac{\partial #1}{\partial #2}}}
\newcommand{\pfi}[2]{\ensuremath{{\partial #1}/{\partial #2}}} 
\newcommand{\mb}[1]{\mathbf{#1}}
\newcommand{\ep}{\varepsilon}
\newcommand{\gb}[1]{\boldsymbol{#1}}

\newcommand{\tr}{\mathrm{tr}\,}

\begin{document}

\DOI{DOI HERE}
\copyrightyear{2025}
\vol{00}
\pubyear{2021}
\access{Advance Access Publication Date: Day Month Year}
\appnotes{Paper}
\copyrightstatement{Published by Oxford University Press on behalf of the Institute of Mathematics and its Applications. All rights reserved.}
\firstpage{1}


\title[Strong anchoring boundary conditions in nematic liquid crystals]{Strong anchoring boundary conditions in nematic liquid crystals:
Higher-order corrections to the Oseen--Frank limit and a revised small-domain theory}

\author{Prabakaran Rajamanickam\ORCID{0000-0003-1240-0362}
\address{\orgdiv{Department of Mathematics}, \orgname{University of Manchester}, \orgaddress{\street{Manchester}, \postcode{M13 9PL},  \country{United Kingdom}}}}

\authormark{Rajamanickam}

\corresp{Corresponding author: \href{email:email-id.com}{prabakaran.rajamanickam@manchester.ac.uk}}

\received{Date}{1}{2026}
\revised{Date}{0}{2025}
\accepted{Date}{0}{2025}


\abstract{Strong anchoring boundary conditions are conventionally modelled by imposing Dirichlet conditions on the order parameter in Landau--de Gennes theory, neglecting the finite surface energy of realistic anchoring. This work revisits the strong anchoring limit for nematic liquid crystals in confined two-dimensional domains. By explicitly retaining a Rapini--Papoular surface energy and adopting a scaling where the extrapolation length $l_{ex}$ is comparable to the coherence length $\xi$, we analyse both the small-domain ($\ep = h/\xi\to 0$; $h$ is the domain size) and Oseen--Frank $(\ep \to \infty$) asymptotic regimes. In the small-domain limit, the leading-order equilibrium solution is given by the average of the boundary data, which can vanish in symmetrically frustrated geometries, leading to isotropic melting. In the large-domain limit, matched asymptotic expansions reveal that surface anchoring introduces an $O(1/\ep)$ correction to the director field, in contrast to the $O(1/\ep^2)$ correction predicted by Dirichlet conditions. The analysis captures the detailed structure of interior and boundary defects, showing that mixed (Robin-type) boundary conditions yield smoother defect cores and more physical predictions than rigid Dirichlet conditions. Numerical solutions for square and circular wells with tangential anchoring illustrate the differences between the two boundary condition treatments, particularly in defect morphology. These results demonstrate that a consistent treatment of anchoring energetics, together with stability considerations, is essential for accurate modelling of nematic equilibria in micro- and nano-scale confined geometries.}


\maketitle

\section{Introduction}
\label{sec:intro}

Nematic liquid crystals are mesomorphic materials characterised by long-range orientational order of rod-like molecules, while retaining liquid-like translational mobility. Their equilibrium configurations arise from a delicate competition between bulk elastic distortions, thermotropic ordering tendencies, interactions with confining boundaries, and external fields~\cite{de1993physics}. In many experimentally relevant settings, such as nematic wells, thin films, and micro or nano patterned devices, equilibrium configurations are strongly influenced and often dictated by surface anchoring conditions. These surface effects play a central role in determining director alignment, defect formation, and the stability of competing nematic states. Consequently, understanding and modelling surface anchoring is essential for the theoretical description of nematic systems in micro and nano scale domains.

From a theoretical perspective, surface anchoring is broadly classified into strong anchoring, weak anchoring, and degenerate anchoring~\cite{mottram2014introduction}. In the strong and weak anchoring cases, surface treatments impose a preferred orientation (easy axis) for the nematic director, with the distinction referring to the magnitude of the energetic penalty associated with deviations from this preferred alignment. Strong anchoring corresponds to surface interactions that dominate over bulk elasticity near the boundary and effectively enforce the preferred orientation. Weak anchoring, by contrast, allows for partial relaxation of the director in response to bulk distortions. Degenerate anchoring prescribes a family of energetically equivalent orientations, which introduces additional freedom and can lead to enhanced defect formation near the boundary.

Strong anchoring conditions are widely employed in analytical and numerical studies of confined nematics, particularly in polygonal and circular domains. In such settings, strong anchoring has been shown to stabilise a rich variety of defect mediated states and symmetry breaking configurations~\cite{majumdar2010landau,lewis2014colloidal,han2020reduced,han2021solution,luo2012multistability,yao2022defect}.

In this work, we focus on strong anchoring within the Landau–de Gennes framework, where it is commonly modelled by imposing Dirichlet boundary conditions on the order parameter. While the traditional quadratic Rapini–Papoular boundary term is standard for non-degenerate alignment, various alternative mathematical descriptions exist in the literature to describe more complex surface anchoring effects. For instance, higher-order expansion models or the fourth-order Fournier–Galatola polynomial free energy density are widely adopted to represent degenerate planar or conically degenerate anchoring settings, where an entire family of spatial orientations is energetically preferred over a single axis~\cite{sen1987landau,fournier2005modeling,mottram2014introduction,golovaty2015dimension}. Nevertheless, for fixed, non-degenerate surface alignments, the quadratic phenomenological model remains highly robust.  The importance of surface anchoring becomes especially evident in geometrically frustrated domains, where the imposed boundary alignment is incompatible across edges or corners. While the Dirichlet boundary conditions are mathematically convenient, they neglect the finite surface energy through which anchoring is realised physically, and may therefore misrepresent defect structures and boundary layers in confined geometries~\cite{ravnik2009landau,majumdar2010landau}. In such cases, anchoring not only enforces global orientational constraints but also drives the formation of boundary defects and localized suppression of nematic order. These effects cannot be captured accurately by models that enforce boundary conditions rigidly, without accounting for the finite energetic cost of anchoring and its interaction with intrinsic nematic length scales such as the coherence length.

Motivated by these considerations, we revisit the notion of strong anchoring within the Landau–de Gennes framework and examine its consequences in confined two-dimensional nematic systems. By retaining surface energy contributions and analysing appropriate asymptotic limits, we demonstrate how anchoring influences both bulk behaviour and defect structures, leading to corrections beyond the classical Oseen--Frank description. We further show that this approach yields physically meaningful predictions in nano scaled systems, where surface effects and intrinsic nematic length scales are necessarily comparable. Our results highlight the essential role of surface anchoring in confined nematics and provide a consistent framework for studying nematic equilibria in small and complex geometries.

\section{Remarks on the strong anchoring condition in three dimensions}

Consider a bounded three-dimensional domain $\Omega$ with piecewise smooth boundary $\partial\Omega$ and characteristic linear dimension $h$, filled with a nematic liquid crystal sample. The order parameter tensor $\mb Q$ belongs to the space of symmetric, traceless $3 \times 3$ matrices. Within the Landau--de Gennes framework, the free energy of the system, supplemented with a Rapini--Papoular surface anchoring energy, is given by
\begin{align}
    F[\mb Q] = \int_\Omega \left[\frac{L}{2}|\nabla\mb Q|^2 + \frac{A}{2}\tr\mb Q^2 -  \frac{B}{3}\tr\mb Q^3 + \frac{C}{4}(\tr\mb Q^2)^2\right]dV + \frac{W}{2}\int_{\partial\Omega}|\mb Q-\mb Q_b|^2 d\Sigma.\label{Feq1}
\end{align}
Here, $A=A(T)$ depends linearly on the temperature $T$,  $L$ and $C$ are positive material constants and both the cubic coefficient $B$ and the anchoring strength $W$ are assumed positive. The tensor $\mb Q_b$ prescribes the preferred surface alignment. The isotropic phase corresponds to $\mb Q = 0$, whereas, the nematic phase with  orientational order corresponds to $\mb Q \neq 0$. In the absence of spatial inhomogeneities and surface effects, the elastic and anchoring contributions in~\eqref{Feq1} may be neglected. In this case, the free energy is globally minimized by
\begin{equation}
    \mb Q_{homo} = \begin{cases}
        \mb 0  & \quad \text{for  } A>B^2/27C,\\
        s_+ (\mb n_+\otimes\mb n_+-\tfrac{1}{3}\mb I)  & \quad \text{for  } A<B^2/27C
    \end{cases}     
\end{equation}
where $s_+=(B+\sqrt{B^2-24AC})/4C$ and $n_+\in \mathbb S^2$ is an arbitrary unit vector. 

In geometrically frustrated domains, however, equilibrium configurations are inherently inhomogeneous, and both bulk and surface energies play an essential role. Strong anchoring is often defined formally by the limit
\begin{equation}
    l_{ex}=\frac{L}{W} \to 0,
\end{equation}
where $l_{ex}$ is the de Gennes--Kleman extrapolation length~\cite{de1993physics}. This length represents the scale over which elastic and surface energies balance. In the formal limit $l_{ex}\to 0$, the surface energy diverges unless $\mb Q=\mb Q_b$ and therefore one enforces the Dirichlet condition $\mb Q=\mb Q_b$ on $\partial\Omega$, while the surface energy itself becomes negligible. Although convenient, this interpretation must be treated with care.

Indeed, when the boundary contains edges or corners, the prescribed boundary tensor $\mb Q_b$ may be geometrically incompatible across intersecting surfaces. For example, tangential anchoring imposed on all faces of a polyhedral domain necessarily leads to discontinuities at vertices and along certain edges~\cite{shi2024multistability}. Such incompatibilities give rise to boundary defects, whose structure cannot be resolved by the extrapolation length alone. Instead, the relevant length scale is the coherence length $\xi$ (on the nematic side), defined by
\begin{equation}
    \xi(T)=\sqrt{\frac{L}{A-2Bs_+/3+2Cs_+^2}},
\end{equation}
which determines the distance over which nematic order is locally suppressed and thus sets the size of defect cores. At the nematic–isotropic transition, $\xi=\sqrt{27LC/B^2}$, while deep in the nematic phase, $\xi\approx \sqrt{L/2|A|}$. By comparing the two length scales introduced above, we can infer that
\begin{itemize}
    \item $l_{ex}\ll \xi$: Surface anchoring dominates down to distances of order $l_{ex}$ and the boundary defects of size $\xi$ are strongly influenced by the surface anchoring. In other words, Dirichlet boundary conditions effectively hold even within the defects.
    \item $l_{ex}\sim \xi$ (distinguished limit): Strong anchoring is satisfied on smooth boundary-data segments but is relaxed within localized defect regions near edges and corners.
    \item $l_{ex}\gg \xi$: Anchoring is effectively weak on scales larger than the defect core. If additionally $l_{ex}\sim h$, weak anchoring prevails across the entire boundary $\partial\Omega$.
\end{itemize}
To place these regimes in context, we note that in practice $l_{ex}\sim 10^{-8}-10^{-5}$m~\cite{ravnik2009landau}, with the strongest achievable anchoring corresponding to a few tens of nanometers. By contrast, the coherence length $\xi\sim 10^{-8}m$~\cite{de1993physics} is typically only a few ångströms and rarely exceeds tens of nanometers even very close to the nematic–isotropic transition\footnote{Strictly speaking, very close to the nematic–isotropic transition, fluctuation effects eventually invalidate the mean-field Landau–de Gennes description.}. For example, using typical material parameters, $L=15pN$, $B=3.7\times 10^5 Nm^{-2}$ and $C=2.4\times 10^5 Nm^{-2}$~\cite{rajamanickam2026colloidal}, one finds $\xi=2.7\times 10^{-8}m$ at the transition point; see also~\cite{gartland2018scalings}. Consequently, the regime $l_{ex}\ll \xi$ is seldom realised in practice. At this stage, it is worth noting that most studies in the literature impose Dirichlet conditions even within localized boundary defects corresponding to this infeasible regime.

Motivated by these considerations, we adopt throughout this work the strong anchoring scaling in which the ratio
\begin{equation}
    \gamma\equiv \frac{\xi}{l_{ex}}\sim O(1) ,   \label{assump}
\end{equation}
thereby allowing for strong anchoring on smooth boundary-data segments while permitting localized relaxation near boundary defects. This scaling is particularly relevant for nanoscale systems, where the domain size $h$ is comparable to $\xi$ or $l_{ex}$. In such regimes, imposing a strict Dirichlet boundary condition would require an unrealistically large surface energy. One would expect partial melting to the isotropic phase when $h\sim \xi$ accompanied by geometric frustration, contrary to the many ordered structures reported in the literature. The exception occurs when the preferred boundary condition does not generate frustration. This observation provides an additional motivation for revisiting the strong anchoring limit and forms a key foundation for the revised small-domain theory developed in this work.

\section{Problem formulation in the reduced two-dimensional Landau--de Gennes framework}

For simplicity and analytical tractability, we focus henceforth on the reduced 2D Landau--de Gennes framework. The reduced framework is suitable for liquid crystal samples confined strongly in the vertical direction such as the nematic walls; top and bottom surfaces are assumed to have degenerate boundary condition, while the lateral surface with strong anchoring in the sense discussed above.

Consider a closed 2D domain $\Omega$ with $\partial\Omega$ being piecewise continuous. In the reduced framework, the order parameter tensor $\mb Q$ belongs to the space of symmetric, traceless $2 \times 2$ matrices, i.e., $\{\mb Q \in \mathbb R^{2 \times 2}\mid \mb Q = \mb Q^T,  \tr\mb Q = 0\}$. The free energy per unit thickness of such a film, along with a Rapini--Papoular-type surface energy for lateral boundaries, can be written as
\begin{align}
    F[\mb Q] = \int_\Omega \left[\frac{L}{2}|\nabla\mb Q|^2 + \frac{A}{2}\tr\mb Q^2 + \frac{C}{4}(\tr\mb Q^2)^2\right]d\mb x + \frac{W}{2}\int_{\partial\Omega}|\mb Q-\mb Q_b|^2 dl\label{Feq}
\end{align}
In the reduced model, $\tr\mb Q^3 = 0$, which rules out biaxiality, and the reduced $\mb Q$-tensor can be written as
\begin{equation}
    \mb Q =  s(\mb n\otimes\mb n- \tfrac{1}{2}\mb I) = \begin{bmatrix}
        q_1 & q_2 \\ q_2 & -q_1
    \end{bmatrix} 
\end{equation}
where $\mb n=(\cos\varphi,\sin\varphi)^T$ is the nematic director, $\varphi$ is the director angle with respect to the $x$-axis and $s$ is the scalar order parameter measuring the degree of molecular alignment along $\mb n$. The relation between $(s,\varphi)$ and $(q_1,q_2)$ are given by
\begin{align}
    q_1 = \tfrac{1}{2}s\cos2\varphi, \qquad q_2 = \tfrac{1}{2}s\sin2\varphi, \qquad  s=2\sqrt{q_1^2+q_2^2}, \qquad \varphi = \frac{1}{2}\tan^{-1}\frac{q_2}{q_1}.  \label{q1q2}
\end{align}
In spatially homogeneous situations without any surface effects, the elastic term in the energy can be neglected, in which case $F$ is globally minimized by
\begin{equation}
    \mb Q_{homo} = \begin{cases}
        \mb 0  & \quad \text{for  } A>0,\\
        s_+ (\mb n_+\otimes\mb n_+-\tfrac{1}{2}\mb I)  & \quad \text{for  } A<0
    \end{cases}     
\end{equation}
where $s_+=\sqrt{-2A/C}$ and $n_+\in \mathbb S^1$ is an arbitrary unit vector in the $xy$-plane.

Rescaling lengths by $h$, the order parameter $\mb Q$ (and $\mb Q_b$) by $s_+$, and the  energy by $s_+^2 L$, the free energy functional reduces to
\begin{align}
      F[\mb Q] = \int_\Omega \left\{\frac{1}{2}|\nabla \mb Q|^2+ \frac{\ep^2}{2}\left[(\tr\mb Q^2)^2-\tr\mb Q^2\right] \right\} d\mb x + \frac{\ep\gamma}{2}\int_{\partial\Omega}|\mb Q-\mb Q_b|^2 dl, \label{freeenergy}
\end{align}
where 
\begin{equation}
    \ep = \frac{h}{\xi} = hs_+\sqrt{\frac{C}{2L}}
\end{equation}
is the ratio of  the domain size $h$ to the 2D reduced coherence length $\xi=\sqrt{L/|A|}$. The surface-energy term now implies that, as the domain size becomes large ($\ep\to \infty$), it grows proportionally indicating a strong anchoring boundary condition, although in very small domains ($\ep\to 0$), it must become weak. The Euler--Lagrange equations for $q_1$ and $q_2$ are given by
\begin{align}
    \nabla^2 q_1 =  \ep^2q_1(4q_1^2+4q_2^2-1), \label{q1}\\  \nabla^2 q_2 =  \ep^2 q_2(4q_1^2+4q_2^2-1)   \label{q2}
\end{align}
which are subject to the boundary conditions
\begin{equation}
    \pfr{q_1}{\nu} = - \ep\gamma (q_1 - q_{1,b}), \quad \pfr{q_2}{\nu} = - \ep\gamma (q_2 - q_{2,b}) \quad \text{on} \quad \partial\Omega, \label{BCs}
\end{equation}
with $\gb\nu$ being the unit outward normal to $\partial\Omega$. The formulation for the associated stability analysis is provided in Appendix A. 

\section{Small-domain limit, $\ep\to 0$}

The solution to the problem~\eqref{q1}-\eqref{BCs} is now described in the asymptotic limit $\ep \to 0$. In the small-domain regime, the solution is unique and admits a regular perturbation expansion of the form
\begin{equation}
    q_1(\mb x) = \sum_{m=0}^\infty \ep^m q_1^{(m)}(\mb x), \qquad q_2(\mb x) = \sum_{m=0}^\infty \ep^m q_2^{(m)}(\mb x).
\end{equation}

The leading-order problem is given by
\begin{align}
    \nabla^2q_1^{(0)}=\nabla^2q_2^{(0)}=0 \quad \text{on}\quad \Omega, \\
     \pfr{q_1^{(0)}}{\nu} =  \pfr{q_2^{(0)}}{\nu} = 0 \quad \text{on} \quad \partial\Omega.
\end{align}
This implies that the leading-order solution is constant throughout the domain. However, this constant is not arbitrary and is instead determined by the solvability condition of the first-order problem. One finds
\begin{align}
    q_1^{(0)} = \frac{1}{|\partial\Omega|}\oint_{\partial\Omega}q_{1,b}\,dl, \qquad q_2^{(0)} = \frac{1}{|\partial\Omega|}\oint_{\partial\Omega}q_{2,b}\,dl, \quad \Rightarrow \quad \mb Q^{(0)} = \frac{1}{|\partial\Omega|}\oint_{\partial\Omega}\mb Q_b\,dl.
\end{align}
Thus, the leading-order solution corresponds to the average of the prescribed boundary data. If we assume $\mb Q_b = s_b(\mb n_b\otimes\mb n_b-\tfrac{1}{2}\mb I_2)$ with $\mb n_b=(\cos\varphi_b,\sin\varphi_b)^T$, then the above solution may be expressed equivalently as
\begin{equation}
    s^{(0)} = \frac{1}{|\partial\Omega|}\left|\oint_{\partial\Omega}s_be^{2i\varphi_b}\,dl\right|, \qquad \varphi^{(0)} = \frac{1}{2}\mathrm{arg}\left(\oint_{\partial\Omega}s_be^{2i\varphi_b}\,dl\right).
\end{equation}
Suppose $s_b=1$, then the complex number $p\equiv \oint_{\partial\Omega} e^{2i\varphi_b}dl$ fully characterises the leading-order solution. This parameter may be compared with the boundary topological degree $d\equiv\tfrac{1}{2\pi}\oint_{\partial\Omega}d\varphi_b(l)$, which plays a central role in the large-domain regime. While $d$ measures a net rotation of the boundary director, $p$ represents a naive average direction of the boundary data. For instance, if $\partial\Omega$ is a circle and $\varphi_b$ corresponds to tangential anchoring, then $d=+1$, whereas $p=0$. In many specialised cases, including regular polygonal domains with highly symmetric geometric frustration imposed by the boundary anchoring, the parameter $p$ may vanish. In such scenarios, $\mb Q^{(0)}=\mb 0$, and the liquid crystal melts to the isotropic state as $\ep\to 0$, with nematic ordering appearing only at order $\ep$.  Previous work by Fang \textit{et al.}~\cite{fang2020surface} considered surface energy effects in the limit $\ep \to 0$, but under the scaling $\gamma\ep \sim 1$, for which a well-ordered structure is obtained at leading order. While it is noted there that $\gamma\ep$ must vanish as $\ep\to0$, the leading-order solution is taken to be an arbitrary constant. By contrast, in the present analysis, the leading-order state is uniquely fixed by the boundary anchoring through the solvability condition.

The first-order problem is given by
\begin{align}
    &\nabla^2 q_1^{(1)}=0, \quad \nabla^2q_2^{(1)}=0 \quad \text{on } \Omega,\\
     &\pfr{q_1^{(1)}}{\nu} = -  \gamma(q_1^{(0)} - q_{1,b}), \quad \pfr{q_2^{(1)}}{\nu} = -  \gamma(q_2^{(0)} - q_{2,b}) \quad \text{on} \quad \partial\Omega,\\
     &\gamma \oint_{\partial\Omega} q_1^{(1)}dl = - q_1^{(0)}(4 q_1^{{(0)}^2} + 4 q_2^{{(0)}^2}-1) |\Omega|, \\   &\gamma\oint_{\partial\Omega} q_2^{(1)}dl = - q_2^{(0)}(4 q_1^{{(0)}^2} + 4 q_2^{{(0)}^2}-1) |\Omega|.
\end{align}
The last two conditions fix the additive constants in $q_1^{(1)}$ and  $q_2^{(1)}$ and arise from the solvability condition of the second-order problem.  The general structure of the higher-order problems for $m\geq 2$ is given by
\begin{align}
     &\nabla^2 q_1^{(m)}= F_1^{(m-2)}, \quad \nabla^2q_2^{(m)}=F_2^{(m-2)} \quad \text{on } \Omega,\\
     &\pfr{q_1^{(m)}}{\nu} = -  \gamma q_1^{(m-1)}, \quad \pfr{q_2^{(m)}}{\nu} = -  \gamma q_2^{(m-1)} \quad \text{on} \quad \partial\Omega,\\
     &\gamma\oint_{\partial\Omega} q_1^{(m)}dl = -\int_{\Omega} F_1^{(m-1)} d\mb x, \\  &\gamma\oint_{\partial\Omega} q_2^{(m)}dl = -\int_{\Omega} F_2^{(m-1)} d\mb x.
\end{align}
where $F_1^{(m)}$ and $F_2^{(m)}$ are the m-th order terms of $q_1 (4 q_1^2+4q_2^2-1)$ and $q_2 (4 q_1^2+4q_2^2-1)$, respectively, i.e.,
\begin{align}
    &F_1^{(m)} = \sum_{k=0}^mq_1^{(k)} P^{(m-k)}  -q_1^{(m)}, \quad F_2^{(m)} = \sum_{k=0}^mq_2^{(k)} P^{(m-k)}  -q_2^{(m)},\\
    &P^{(m)} = 4\sum_{k=0}^{m}\left[q_1^{(k)}q_1^{(m-k)} + q_2^{(k)}q_2^{(m-k)}\right].
\end{align}

In summary, in the small-domain limit the problem reduces to a hierarchy of Poisson equations with Neumann boundary conditions, supplemented by integral constraints that uniquely determine the additive constants at each order.

\section{Oseen--Frank limit, $\ep\to \infty$}

We now consider the large-domain or Oseen--Frank limit $\ep\to\infty$, which is the most practically relevant limit. In this limit, the solution is generally multi-valued and the limit represents a singular perturbation problem. Consequently, the solution structure comprises an outer (bulk) region and multiple inner regions. The outer region corresponds to defect-free domains, while the inner regions describe interior or boundary defects. Within the Landau--de Gennes framework, the defects may appear only at a finite number of isolated singular points $\{x_k\}$ with charges $m_k$, $k=1,2,\dots,N$, as shown in~\cite{bethuel1994ginzburg,majumdar2010landau}. For a given geometry and boundary conditions, there exists a minimal choice of $N$ and $|m_k|$ that represents the ground state  or global minimiser, determined by the trade-off between geometric frustration and defect energetics. Table~\ref{tab:corner} summarises the defect structure for several canonical geometries subject to tangential anchoring. Excited states (local minimizers) will have larger $N$ or $|m_k|$ or both.

\begin{table*}[h]
\centering
\footnotesize
\caption{Defect characterisation for the ground state subject to tangential anchoring.} \label{tab:corner}
  \begin{tabular}{|c|c|c|} \hline
  Geometry & $N$ & Defects character   \\ \hline
   Equilateral triangle~\cite{rajamanickam2026nematic} & $4$ & three $+1$-vertex defects \& one $-\tfrac{1}{2}$-interior defect  \\
   Rectangle~\cite{luo2012multistability} & $4$ & two $+1$-vertex defects \& two $-1$-vertex defects \\
  Circle &   $2$ & two $+\tfrac{1}{2}$-interior defects \\   
  Pentagon~\cite{han2020reduced} &  $5$ & two $+1$-vertex defects \& three $-\frac{2}{3}$-vertex defects \\
   Hexagon~\cite{han2020reduced} & $6$ & two $+1$-vertex defects \& four $-\frac{1}{2}$-vertex defects \\ 
   Isosceles triangle with apex angle $<41^\circ$~\cite{rajamanickam2026nematic} &  $3$ & two $+1$-vertex defects \& one $\frac{2\alpha}{2\alpha-\pi}$-vertex defect\\
  Isosceles triangle with apex angle $>72^\circ$~\cite{rajamanickam2026nematic} &  $3$ & two $+1$-vertex defects \& one $\frac{2\alpha-\pi}{2\alpha}$-vertex defect  \\\hline
  \end{tabular}
\end{table*}

In the Oseen--Frank limit, it is natural to assume $s_b=1$ so that
\begin{equation}
    \mb Q_b = \mb n_b\otimes\mb n_b- \tfrac{1}{2}\mb I, \qquad \mb n_b=\begin{bmatrix}
        \cos\varphi_b \\ \sin\varphi_b
    \end{bmatrix}.
\end{equation}
 To facilitate the analysis for the Oseen--Frank limit, it is convenient to adopt the $(s,\varphi)$-formulation, in place of the $(q_1,q_2)$-formulation. The Euler--Lagrange equations for $s$ and $\varphi$ take the form
\begin{align}
    \nabla^2 s - 4s|\nabla\varphi|^2 &=\ep^2 s(s^2-1), \\ \nabla\cdot(s^2 \nabla \varphi)&=0.
\end{align}
With $s_b=1$, the boundary conditions are given by
\begin{equation}
      \pfr{s}{\nu} = -\ep\gamma[s-\cos(2\varphi-2\varphi_b)], \qquad   2s\pfr{\varphi}{\nu} = -\ep\gamma \sin(2\varphi-2\varphi_b) \quad \text{on} \quad \partial\Omega. \label{sphibc}
\end{equation}
Below, we provide a prescription of the solution in the asymptotic limit $\ep\to \infty$ using the language of method of matched asymptotic expansions. This construction applies to any admissible solution of the leading-order problem. 

\subsection{Outer solution} \label{sec:outeroseenfrank}
For the outer defect-free bulk region, we can introduce the perturbation series
\begin{align}
    s = \sum_{m=0}^\infty \ep^{-m} s_m, \qquad \varphi = \sum_{m=0}^\infty \ep^{-m} \varphi_m.
\end{align}
Substituting these series expansions into the boundary conditions~\eqref{sphibc} at leading order yields $s_0=\cos(2\varphi_0-2\varphi_b)$ and $\sin(2\varphi_0-2\varphi_b)=0$. The second relation implies $\varphi_0 = \varphi_b \pmod{\pi/2}$, where the modulo factor can be dropped by appropriately shifting the reference anchoring angle $\varphi_b$. Choosing the standard positive order parameter branch $s_0=1$ from the first relation then enforces the Dirichlet boundary conditions $\varphi_0=\varphi_b$ and $s_0=1$ on $\partial\Omega$. Working out the bulk equations similarly, the leading-order problem reduces to
\begin{align}
    s_0=1, \quad \nabla^2 \varphi_0=0 \quad &\text{in} \quad \Omega,  \label{OFleading}\\
    s_0=1, \quad \varphi_0=\varphi_b  \quad &\text{on} \quad \partial\Omega.  \label{OFleadingBC}
\end{align}
At first order, the problem simplifies to 
\begin{align}
    s_1=0, \quad \nabla^2 \varphi_1=0 \quad &\text{in} \quad \Omega, \label{firstorderouter1}\\
    s_1=0, \quad \varphi_1=-\frac{1}{\gamma}\pfr{\varphi_0}{\nu}  \quad &\text{on} \quad \partial\Omega. \label{firstorderouter2}
\end{align}
At second order, the problem simplifies to 
\begin{align}
    s_2=-2|\nabla\varphi_0|^2, \quad \nabla^2 \varphi_2=4\nabla\cdot(|\nabla\varphi_0|^2\nabla\varphi_0) \quad &\text{in} \quad \Omega, \label{s2phi2}\\
    s_2=-\frac{2}{\gamma^2}\left(\pfr{\varphi_0}{\nu}\right)^2, \quad \varphi_2=-\frac{1}{\gamma}\pfr{\varphi_1}{\nu}  \quad &\text{on} \quad \partial\Omega. \label{s2phi2bc}
\end{align}
The mismatch between bulk $s_2$ and the boundary $s_2$ arises from $\gamma^2\neq 1$ and $\pfi{\varphi_0}{\tau}\neq 0$, indicating a weak boundary layer in $s$. Formally, above problems are defined on $\Omega\backslash\{x_k\}$, but in the matched asymptotic expansion framework, the singular points are accounted for through the inner expansions, and the composite expansion handles overlaps automatically.

The first-order correction exists only because $\gamma\sim O(1)$. In studies where surface energy is neglected $(\gamma \to \infty)$, $\varphi_1=0$ and the first correction vanishes. Properly accounting for surface energy introduces an $O(1/\ep)$ correction to the director field, in contrast to the $O(1/\ep^2)$ prediction from purely Dirichlet boundary conditions~\cite{nguyen2013refined,di2020landau}. The bulk scalar order parameter, however, satisfies
\begin{equation}
    s = 1 -\frac{2}{\ep^{2}} |\nabla\varphi_0|^2 - \frac{4}{\ep^{3}} \nabla\varphi_0\cdot \nabla\varphi_1  -  \cdots.   \label{outers}
\end{equation}
The director field $\mb n=(\cos\varphi,\sin\varphi)^T$ admits the expansion
\begin{equation}
    \mb n = \mb n_0 + \ep^{-1} \varphi_1 \mb t_0 + \ep^{-2}(\varphi_2\mb t_0 - \tfrac{1}{2}\varphi_1^2\mb n_0)+\cdots
\end{equation}
where
\begin{align}
    \mb n_0= \begin{bmatrix}
        \cos\varphi_0 \\ \sin\varphi_0
    \end{bmatrix},\qquad \mb t_0 = \begin{bmatrix}
        -\sin\varphi_0 \\ \cos\varphi_0
    \end{bmatrix}.
\end{align}
The $O(1/\ep)$ term represents partial relaxation of the director along $\mb t_0$ in response to leading-order bulk elastic stresses. This correction arises from retaining the surface energy under the strong anchoring scaling $\gamma=\xi/l_{ex}\sim 1$. The expansion for the $\mb Q$-tensor is given by
\begin{equation}
    \mb Q= \mb Q_0 + \ep^{-1}\varphi_1\mb T_0 +\ep^{-2}[\varphi_2\mb T_0 - (|\nabla\mb Q_0|^2+2\varphi_1^2)\mb Q_0] +  \cdots \label{Qexpansion}
\end{equation}
where
\begin{align}
    \mb Q_0 &= \left. \pfr{\mb Q}{s} \right|_{s=1,\varphi=\varphi_0}= \mb n_0\otimes\mb n_0 -\tfrac{1}{2}\mb I_2, \\ \mb T_0 & = \left. \pfr{\mb Q}{\varphi} \right|_{s=1,\varphi=\varphi_0}= \mb n_0\otimes\mb t_0+\mb t_0\otimes\mb n_0 
\end{align}
with $\mb Q_0:\mb T_0=0$. The $\ep^{-2}\mb Q_0$ in~\eqref{Qexpansion} corresponds to $\pfi{^2\mb Q}{\varphi^2}=-4\mb Q$, whereas the $\ep^{-2}\mb T_0$ corresponds to the mixed derivative $\pfi{^2\mb Q}{s\partial\varphi}=\mb T$, evaluated at $s=1$ and $\varphi=\varphi_0$. In fact, the exact closed-form expansion for the $\mb Q(s,\varphi)$ about  $s=1$ and $\varphi=\varphi_0$ is 
\begin{equation}
    \mb Q(s,\varphi) = (1+\delta s) \left[\cos(2\delta\varphi)\mb Q_0 + \tfrac{1}{2}\sin(2\delta\varphi) \mb T_0\right], \quad \delta s= s-1, \quad \delta\varphi = \varphi-\varphi_0
\end{equation}
from which the asymptotic series in $\ep^{-1}$ follows upon substituting $\delta s= \ep^{-2}s_2 + \cdots$ and $\delta\varphi = \ep^{-1}\varphi_1 + \ep^{-2}\varphi_2 + \cdots$.

In the limit $\gamma\to \infty$ (i.e., Dirichlet boundary conditions), $\varphi_1=0$. For this case, $O(\ep^{-2})$ term proportional to $\mb Q_0$ was obtained earlier by Di Fratta \textit{et. al.}~\cite[Section 4]{di2020landau} as an $O(\ep^{-2})$ correction affecting only the scalar order parameter. Our expansion provides the full $O(\varepsilon^{-2})$ correction in $\mathbf{Q}$, including the component $\varphi_2\mathbf{T}_0$ orthogonal to $\mathbf{Q}_0$, which corresponds to a second-order correction in the director field not previously computed. 

\begin{figure}[h]
\centering
\includegraphics[width=0.4\textwidth]{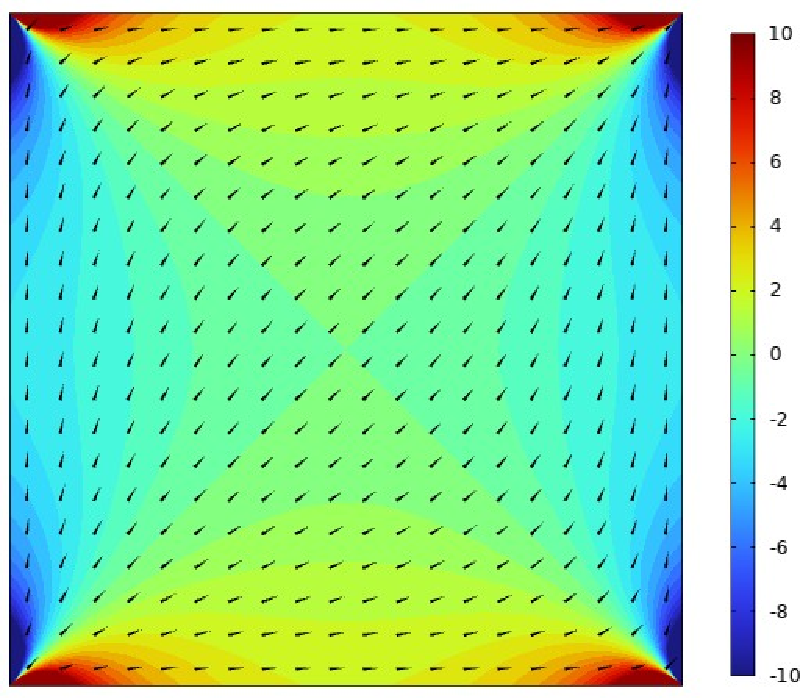}
\caption{Contours of $\varphi_1$ for the diagonal state in a square well, calculated with $\gamma=1$. The director field pertains to $\mb n_0$. The colour contours are limited to the range $[-10,10]$ in order to visualise $\varphi_1$ in the bulk and avoid its corner singularities.} 
\label{fig:oseenfrank}
\end{figure}

Figure~\ref{fig:oseenfrank} shows $\varphi_1$ for the diagonal state in a square well with tangential anchoring. Both $\varphi_1$ and higher-order corrections such as $\varphi_2$ are singular at the corners. To analyse the behaviour near a corner of charge $m$, introduce local polar coordinates $(r,\theta)$ centred at the corner. Since $\nabla\varphi_0 = (m/r)\mathbf{e}_\theta$ near the corner, the boundary condition $\varphi_1 = \partial\varphi_0/\partial\nu$ takes opposite signs on the two adjacent edges, leading to $\varphi_1 \sim O(1/r)$ as $r\to 0$. Similarly, $\varphi_2 \sim O(1/r^2)$. In general, near a corner one finds the asymptotic scaling $\varphi_k \sim O(1/r^k)$ as $r\to 0$. These singularities must be resolved in an inner region where $\varepsilon^{-k}\varphi_k$ becomes $O(1)$, which occurs when $r \sim 1/\varepsilon$.

For interior defects the situation is different. Again take local polar coordinates $(r,\theta)$ centred at an interior defect of charge $m$. Here $\varphi_1$ remains bounded, $\varphi_1 \sim O(1)$ as $r\to 0$. Moreover, because $\varphi_0 = m\theta + \text{const.}$, we have $\nabla^2\varphi_2 \approx 0$ near $r=0$, and consequently $\varphi_2$ is also bounded at the defect centre.

In certain symmetric cases, $\varphi_1$ may vanish identically even for $\gamma = O(1)$. For instance, in a circular well with tangential anchoring, one typically has $\partial\varphi_0/\partial\nu = 0$ on $\partial\Omega$, so Dirichlet boundary conditions are adequate in the large‑domain limit. Broadly speaking, the influence of surface anchoring on the defect structure and director configuration becomes more pronounced as the geometric frustration imposed by the boundary increases.

\subsection{Structure of an interior defect}

Consider an interior defect with charge $m$. Let us define a local coordinate system $(r,\theta)$ centred at the defect. Near the defect, the leading-order outer solution behaves as
\begin{equation}
    s_0=1, \quad \varphi_0 = m \theta + \text{const.,}
\end{equation}
where the constant describes the orientation of the defect. The inner region or defect core, has size $r\sim 1/\ep$ and corresponds to a region where the scalar order parameter vanishes at the centre. To describe the defect structure, we introduce the stretched coordinates
\begin{equation}
    \eta = \ep r, \qquad \theta=\theta
\end{equation}
and write the asymptotic series for the inner region as
\begin{align}
    s(\eta,\theta)= S_0 + \ep^{-1} S_1 + \ep^{-2} S_2 + \cdots \\
    \varphi(\eta,\theta)= \Phi_0 + \ep^{-1} \Phi_1 + \ep^{-2} \Phi_2 + \cdots.
\end{align}

It is easy to verify that the leading-order solution is given by
\begin{align}
    S_0 = S_0(\eta), \qquad \Phi_0 = m \theta + \text{const.}, 
\end{align}
where $S_0$, an even function of $\eta$, satisfies the radial problem
\begin{align}
    \frac{1}{\eta}\frac{d}{d\eta}\left(\eta\frac{dS_0}{d\eta}\right) - \frac{4m^2}{\eta^2}S_0 = S_0^3-S_0, \quad S_0(0)=0, \quad S_0(\infty)=1.
\end{align}
\begin{figure}[h]
\centering
\includegraphics[width=0.5\textwidth]{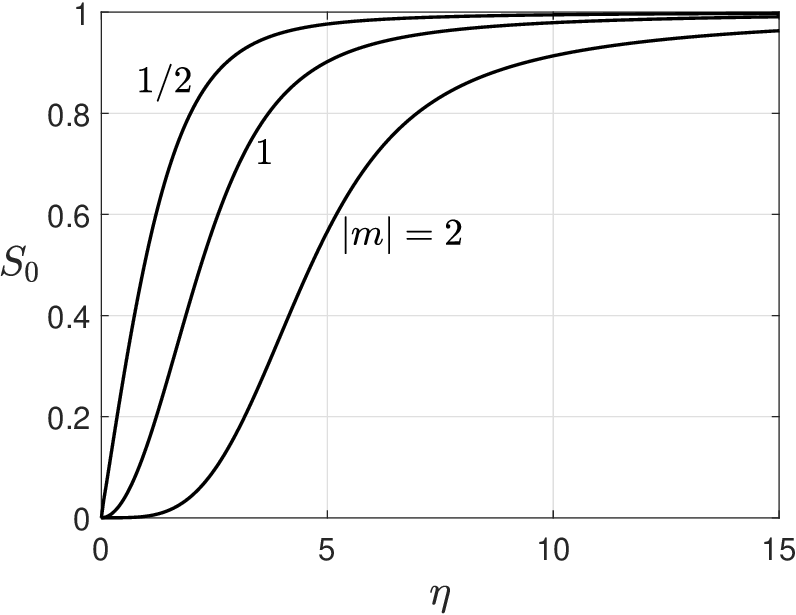}
\caption{The leading-order scalar order parameter $S_0(\eta)$ for three values of the defect charge $|m|$.} 
\label{fig:S0}
\end{figure}
 The function $S_0(\eta)$ can be computed numerically and is illustrated in Fig.~\ref{fig:S0}, for three values of $|m|$. Its asymptotic behaviour is found to be
\begin{align}
    S_0 \sim \eta^{2|m|} \quad \text{as} \quad \eta\to0, \quad \text{and} \quad
    S_0\approx 1 - \frac{2m^2}{\eta^2}  \quad \text{as} \quad \eta\to\infty. \label{radial}
\end{align}
The second asymptotic behaviour, expressed in terms of the outer variables, implies that $s_0 \approx 1 - 2m^2/\ep^2r^2$ as $r\to 0$, which matches exactly with~\eqref{outers} since $\nabla\varphi_0= (m/r)\mb e_\theta$. It is worth noting that the defect structures emerging in the reduced Landau--de Gennes framework, which is applicable to strongly confined systems, differ from other types of reduced models, such as planar and axisymmetric configurations embedded in three dimensions~\cite{schopohl1987defect,penzenstadler1989fine}.

The first-order inner problem satisfies the following coupled linear PDEs
\begin{align}
    \nabla_\eta^2 S_1 - \left(\frac{4m^2}{\eta^2}+3S_0^2-1\right) S_1 &=\frac{8 m}{\eta^2}S_0\pfr{\Phi_1}{\theta}, \label{firstorderinner1} \\
     \nabla_\eta\cdot(S_0^2 \nabla_\eta\Phi_1)  &=-\frac{2 m}{\eta^2}\pfr{(S_0S_1)}{\theta}\label{firstorderinner2}
\end{align}
where $\nabla_\eta$ is the gradient operator for the inner coordinates $(\eta,\theta)$. The above PDEs are subject to the boundary conditions
\begin{align}
    \eta=0: &\quad S_1=0, \quad \Phi_1 \text{ is bounded}, \label{firstorderinner3}\\
    \eta\to \infty: &\quad S_1\to 0, \quad \Phi_1\to F(\theta), \quad F(\theta)\equiv \varphi_1(r\to 0,\theta),   \label{firstorderinner4} \\
    S_1(\eta,\theta)&= S_1(\eta,\theta+2\pi), \quad  \Phi_1(\eta,\theta)= \Phi_1(\eta,\theta+2\pi).\label{firstorderinner5}
\end{align}
The $2\pi$-periodic function $F(\theta)$, determined from the solution of the first-order outer problem~\eqref{firstorderouter1}-\eqref{firstorderouter2} drives the inner problem. Its angular variation encodes the global asymmetry of the outer solution. In highly symmetric cases, such as a circular well with a central $+1$-defect, $\varphi_1=0$ and hence $F=0$. In the general case where $F$ varies with $\theta$, $(S_1,\Phi_1)$ become nontrivial, producing asymmetric defect cores at order $1/\ep$. This is a novel result, as it demonstrates that surface energy contributions generate asymmetry at first order, which would otherwise appear only at order $1/\ep^2$ if surface energy were neglected.  The solution to this first-order inner problem can be reduced to a system of uncoupled ordinary differential equations for each mode $n \in \mathbb{Z}$ by substituting the Fourier series representations
\begin{equation}
    F=\sum_{n\in\mathbb Z} \hat F_n e^{in\theta}, \qquad S_1=\sum_{n\in\mathbb Z} \hat S_{1,n}(\eta) e^{in\theta}, \qquad \Phi_1=\sum_{n\in\mathbb Z} \hat\Phi_{1,n}(\eta) e^{in\theta}
\end{equation}
into equations~\eqref{firstorderinner1}--\eqref{firstorderinner2}. Since $F$, $S_1$, and $\Phi_1$ are real-valued functions, their complex Fourier coefficients satisfy the Hermitian conditions $\hat F_{-n} = \hat F_n^*$, $\hat S_{1,-n} = \hat S_{1,n}^*$, and $\hat \Phi_{1,-n} = \hat \Phi_{1,n}^*$. Collecting coefficients for each $e^{in\theta}$ mode yields the following linear system of second-order ODEs for the radial amplitudes $\hat S_{1,n}(\eta)$ and $\hat \Phi_{1,n}(\eta)$:
\begin{align}
   \frac{1}{\eta}\frac{d}{d\eta}\left(\eta\frac{d\hat S_{1,n}}{d\eta}\right)   - \left(\frac{4m^2+n^2}{\eta^2} + 3S_0^2 - 1\right)\hat S_{1,n} &= \frac{8imn}{\eta^2}S_0\hat \Phi_{1,n}, \label{inner_ode1} \\
    \frac{1}{\eta}\frac{d}{d\eta}\left(\eta S_0^2 \frac{d\hat \Phi_{1,n}}{d\eta}\right)  - \frac{n^2}{\eta^2}S_0^2\hat \Phi_{1,n} &= -\frac{2imn}{\eta^2}S_0 \hat S_{1,n}. \label{inner_ode2}
\end{align}
Utilizing the far-field matching and regularity criteria outlined in~\eqref{firstorderinner3}--\eqref{firstorderinner4}, the boundary conditions for the system of ODEs~\eqref{inner_ode1}--\eqref{inner_ode2} for each mode $n$ reduce to
\begin{align}
    \text{at } \qquad \eta = 0: & \quad \hat S_{1,n} = 0, \quad \hat \Phi_{1,n} \text{ is bounded}, \label{ode_bc1} \\
    \text{as } \qquad \eta \to \infty: & \quad \hat S_{1,n} \to 0, \quad \hat \Phi_{1,n} \to \hat F_n. \label{ode_bc2}
\end{align}
Thus, the complex far-field Fourier components $\hat F_n$ of the outer director correction directly dictate the boundary conditions for the inner problem at infinity, uniquely driving the asymmetric core profile. Notably, if the global outer solution is such that the far-field data vanishes ($\hat F_n = 0$ for all $n \in \mathbb{Z}$), the homogeneous boundary conditions at infinity imply that the system admits only the trivial solution, $\hat S_{1,n}(\eta) = \hat \Phi_{1,n}(\eta) = 0$. In such highly symmetric cases, core asymmetries are entirely suppressed at this order, and the defect core remains perfectly symmetric up to $O(1/\epsilon)$.

\subsection{Structure of a boundary defect}

Boundary defects arise either from discontinuities in the boundary data along a smooth surface or from non-smooth boundary features such as corners and cusps. For simplicity, we consider a corner formed by the intersection of two tangent lines $\theta=\pm \alpha$ with $2\alpha\in (0,2\pi)$. 

Near the corner, the leading-order outer solution behaves as
\begin{equation}
    r\to 0: \quad s_0=1, \quad \varphi_0 = m\theta + \text{const}.
\end{equation}
Unlike interior defects, the charge $m$ of a corner defect need not be an integer or half-integer. For example, consider two basic corner types,
\begin{align}
    \text{Splay corner:} & \quad \varphi_b = +\alpha \quad \text{on} \quad \theta = + \alpha, \quad \text{and}\quad \varphi_b = -\alpha \quad \text{on} \quad \theta = - \alpha,\\
    \text{Bend corner:} & \quad \varphi_b = +\alpha \quad \text{on} \quad \theta = + \alpha ,\quad \text{and}\quad \varphi_b = \pi-\alpha \quad \text{on} \quad \theta = - \alpha.
\end{align}
Then $m=+1$ for the splay corner and $m=1-\pi/2\alpha$ for the bend corner. 

As with interior defects, we introduce stretched coordinates $(\eta,\theta)$ and an inner asymptotic expansion to describe the corner structure. Unlike the interior defect, the leading-order problem for the corner does not decouple. We obtain
\begin{align}
    \nabla_\eta^2S_0^2 - 4 S_0|\nabla\Phi_0|^2&=S_0^3-S_0, \\ \nabla_\eta\cdot(S_0^2\nabla_\eta\Phi_0)&=0
\end{align}
together with the boundary conditions
\begin{align}
    &\frac{1}{\gamma\eta}\pfr{S_0}{\theta}=S_0-\cos(2\Phi_0-2\varphi_b),\quad \text{on} \quad \theta=\pm\alpha\\ & \frac{2S_0}{\gamma\eta}\pfr{\Phi_0}{\theta}= \sin(2\Phi_0-2\varphi_b) \quad \text{on} \quad \theta=\pm\alpha,\\
    &S_0\to 1, \qquad \Phi_0\to m\theta+\text{const.} \quad \text{as} \quad \eta \to \infty,\\
    &S_0\to 0, \qquad \Phi_0 \text{ is bounded} \quad \text{as} \quad \eta \to 0.
\end{align}
The solution of this problem captures the relaxation of the nearly Dirichlet boundary condition in the vicinity of the corner, describing the detailed structure of the boundary defect and its transition to the bulk director field.

\section{Numerical results with tangential anchoring}

In this section, we present numerical solutions of the problem~\eqref{q1}-\eqref{BCs} for few illustrative domains subjected to tangential anchoring, with $s_b=2(q_{1,b}^2+q_{2,b}^2)^{\frac{1}{2}}=1$. In all  numerical computations, we set $\gamma=1$. For comparison, we also consider the $\ep$-independent Dirichlet boundary condition
\begin{equation}
   q_1 = q_{1,b}, \qquad q_2 = q_{2,b} \quad \text{on} \quad \partial\Omega. \label{BCDcs}
\end{equation}
The governing nonlinear equations are solved numerically using the Finite Element Method implemented in COMSOL Multiphysics,  utilizing quadratic Lagrange elements on a highly refined unstructured triangular mesh with a maximum element size of $1/500$ to ensure mesh-independent resolution of the defect core fine structures.

\subsection{Square well}

Consider a square well $\Omega:=\{x\in[0,1],y\in[0,1]\}$ with tangential anchoring  so that
\begin{align}
    q_{1,b} = +\tfrac{1}{2}, \quad q_{2,b}=0 &\quad \text{on horizontal edges},\\
    q_{1,b} = -\tfrac{1}{2}, \quad q_{2,b}=0 &\quad \text{on vertical edges}.
\end{align}

\begin{figure}[h]
\centering
\includegraphics[width=0.45\textwidth]{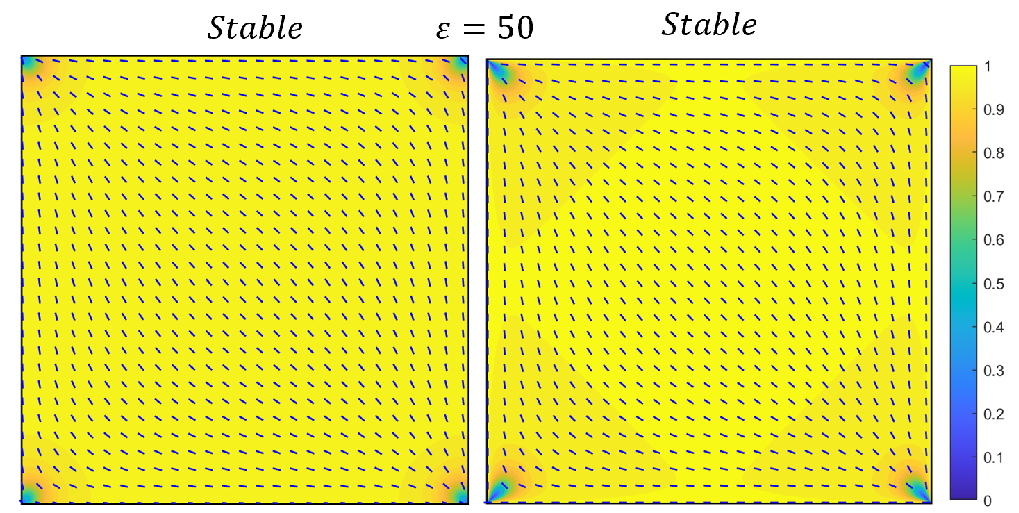}
\includegraphics[width=0.45\textwidth]{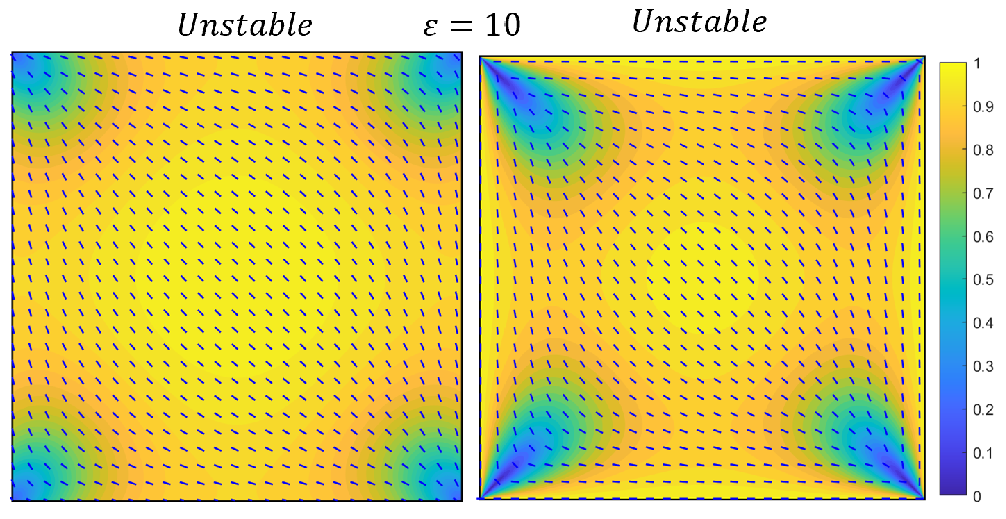}
\includegraphics[width=0.45\textwidth]{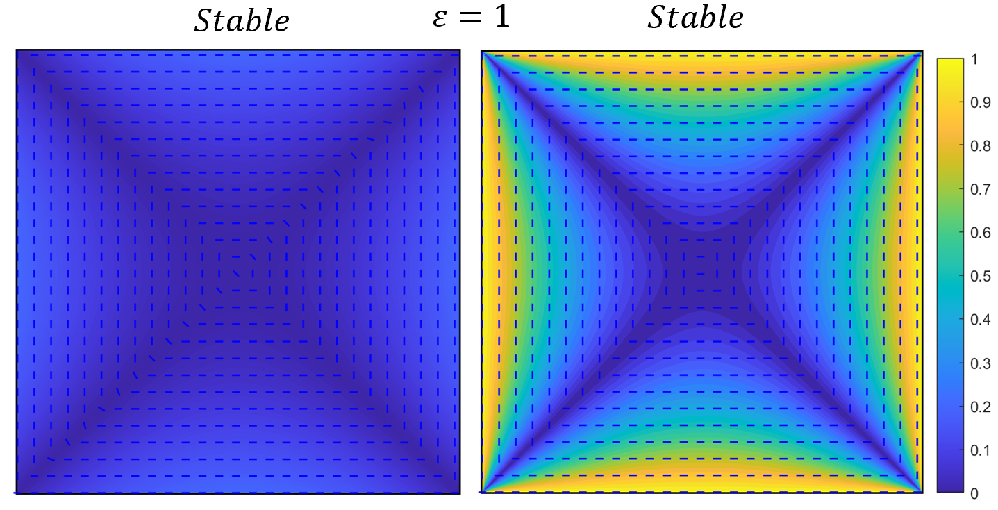}
\includegraphics[width=0.45\textwidth]{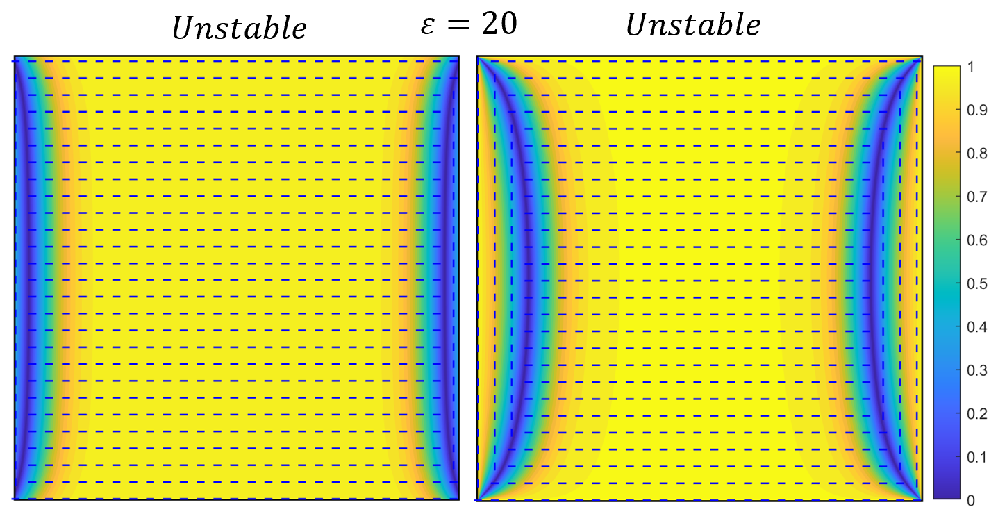}
\caption{Equilibria in a square well. In each pair of figures, the left plots correspond to our mixed boundary conditions~\eqref{BCs}, whereas the right ones correspond to the Dirichlet boundary conditions~\eqref{BCDcs}. The colour contour represent the scalar order parameter $s$. The top row corresponds to the diagonal (D) states, the first pair in the bottom row to the cross-defect (X) states, and the last pair to the boundary-defect (BD) states.} 
\label{fig:square}
\end{figure}

Representative numerical results obtained with the mixed boundary conditions~\eqref{BCs} and Dirichlet boundary conditions~\eqref{BCDcs} are shown in Fig.~\ref{fig:square}. The top rows correspond to the diagonal (D) states for $\ep=50$ and $\ep=10$, which feature two splay and two bend corners. One immediate observation concerns the structure of the corner defect cores. The mixed boundary conditions reveal a smooth and gradual relaxation of the scalar order parameter within the defect core, similar to the behaviour reported in~\cite{luo2012multistability}. In contrast, the Dirichlet boundary conditions produce highly distorted defect cores. The first pair in the bottom row, corresponding to $\ep=1$ (representative of the small-domain limit), shows the diagonal-cross (X) or Well-Order Reconstruction Solution (WORS)~\cite{canevari2020well}. In this case, the mixed boundary condition solution is almost melted to the isotropic phase, with the scalar order parameter vanishingly small. Specifically, as $\ep \ll 1$, we have
\begin{align}
    q_1 &= \frac{\ep\gamma}{2}(y^2-x^2+x-y)  + O(\ep^2),\\
    q_2 &= 0 \quad \text{to all orders in }\ep
\end{align}
so that the diagonal lines $y=x$ and $y=1-x$ are defects and $s=2q_1\sim \ep\gamma$. 
By contrast, the Dirichlet boundary condition enforces an ordered structure even in the small-domain regime. The last pair depicts the boundary-defect (BD) state for $\ep=20$, where two line defects appear along opposite edges. Since these defects lie close to the boundaries, their structure is strongly influenced by the choice of boundary conditions, with mixed conditions producing smoother defect cores than the Dirichlet prescription.

 \begin{figure}[h]
\centering
\includegraphics[width=0.45\textwidth]{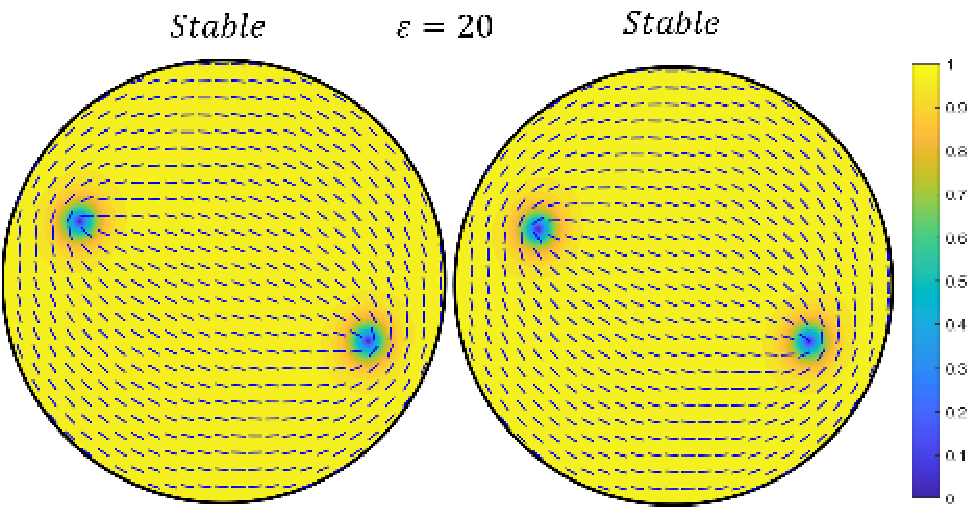}
\includegraphics[width=0.45\textwidth]{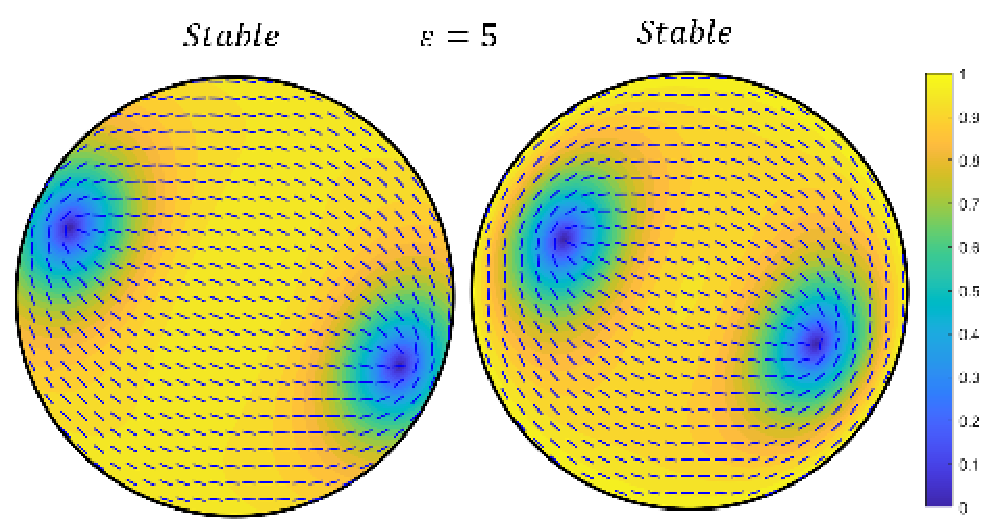}
\includegraphics[width=0.45\textwidth]{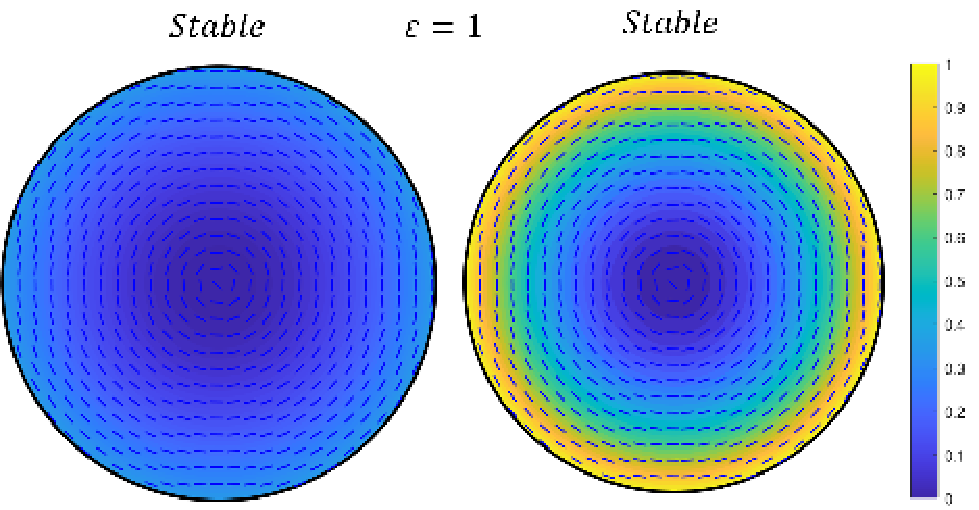}
\includegraphics[width=0.45\textwidth]{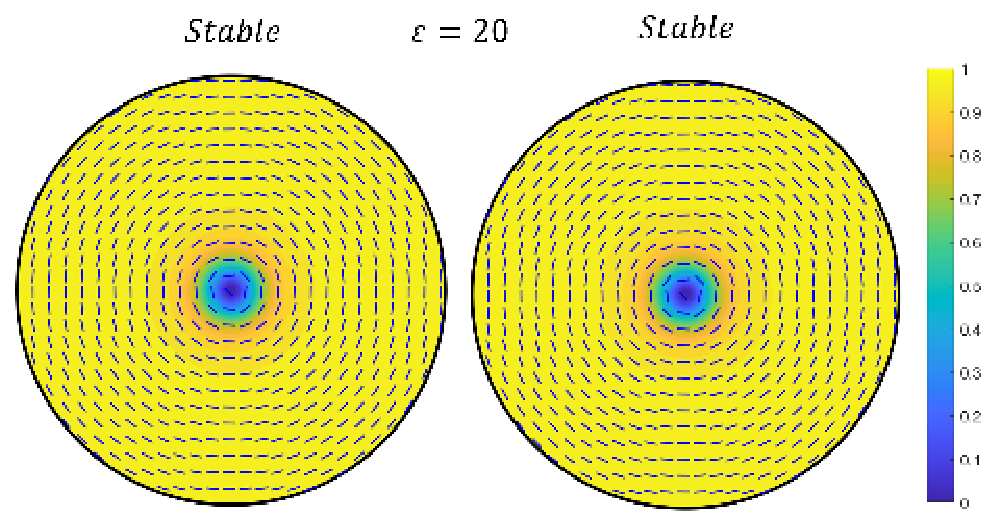}
\caption{Equilibria in a circular well. In each pair of figures, the left plots correspond to our mixed boundary conditions~\eqref{BCs}, whereas the right ones correspond to the Dirichlet boundary conditions~\eqref{BCDcs}. The colour contour represent the scalar order parameter $s$. The top row corresponds to the state with two $+\tfrac{1}{2}$-defects, whereas the bottom row corresponds to the  state with one $+1$-defect (a vortex or ring solution).} 
\label{fig:circle}
\end{figure}

\begin{figure}[h]
\centering
\includegraphics[width=0.7\textwidth]{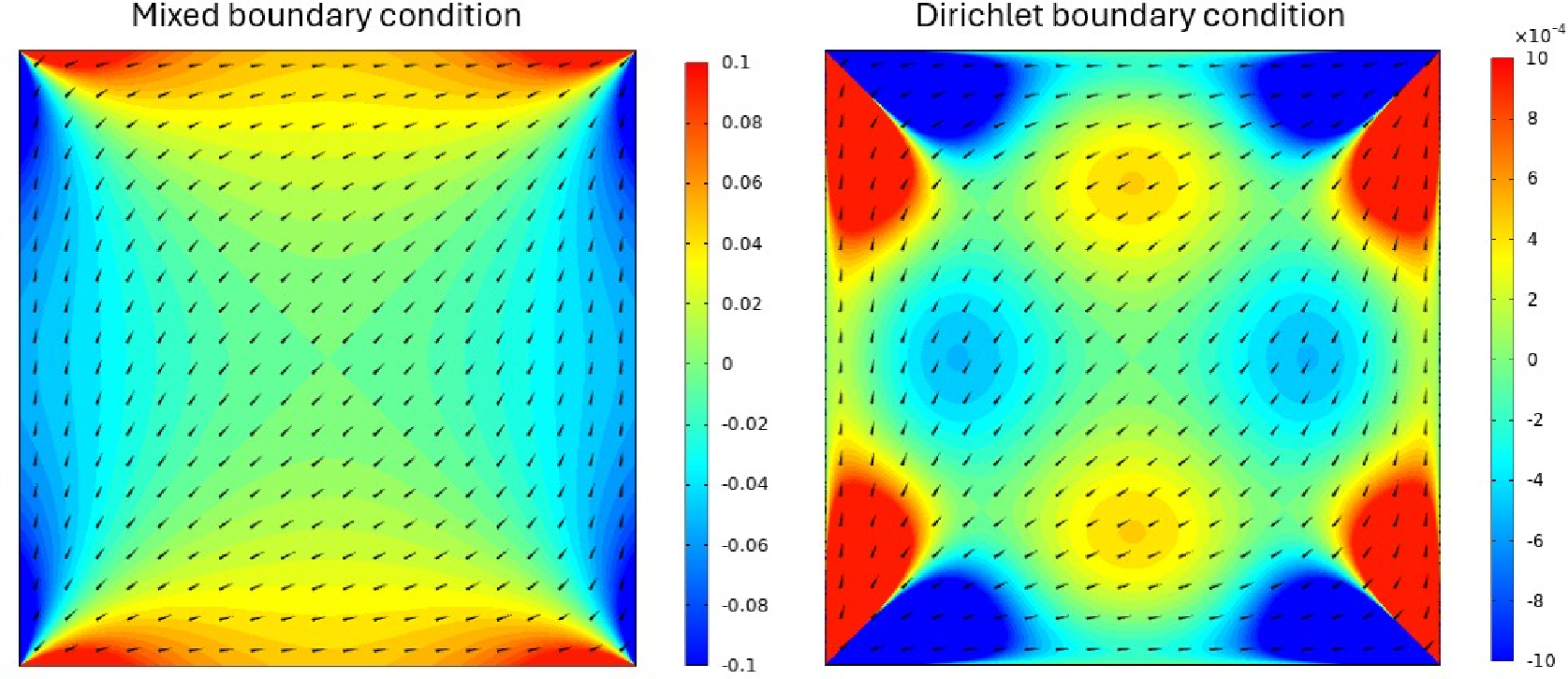}\vspace{0.3cm}
\includegraphics[width=0.35\textwidth]{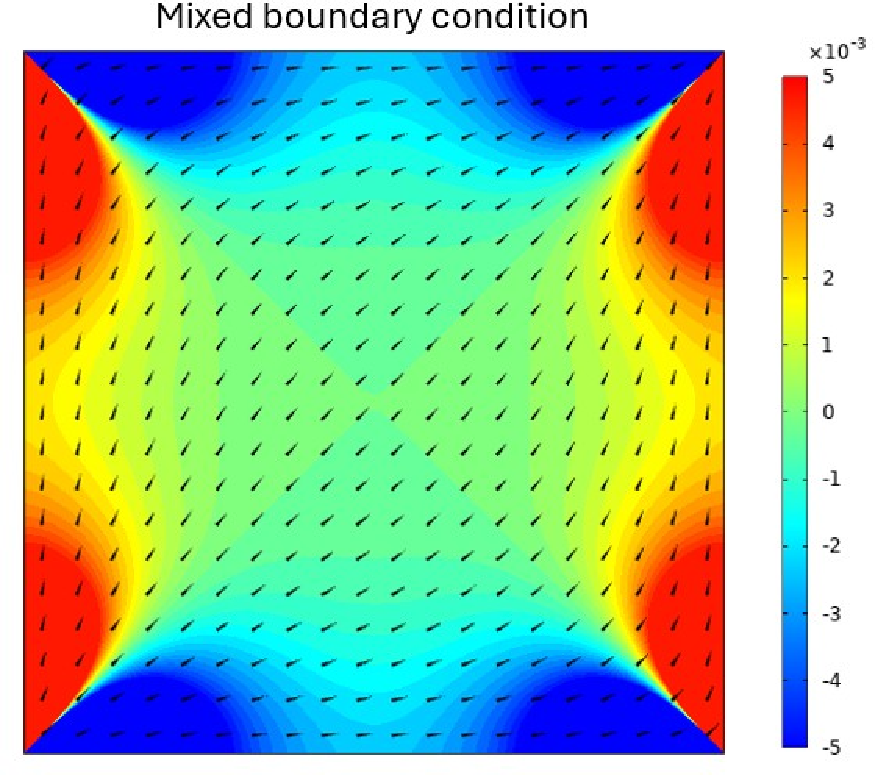}
\caption{Contours of the director angle corrections in a square well at $\ep = 50$. Top-left panel: The first-order deviation $\Delta \varphi = \varphi - \varphi_{\text{OF},0}$ for the finite surface anchoring model ($\gamma = 1$) restricted to $[-0.1, 0.1]$, revealing a dominant $O(1/\ep)$ bulk pattern ($\varphi_1$) similar to the one observed in Fig.~\ref{fig:oseenfrank}. Top-right panel: The deviation $\Delta \varphi = \varphi - \varphi_{\text{OF},0}$ for the rigid Dirichlet model restricted to $[-0.001, 0.001]$, isolating the higher-order $O(1/\ep^2)$ bulk pattern ($\varphi_2$). Bottom panel: The second-order residual field $\varphi - \varphi_{\text{OF},0} - \ep^{-1}\varphi_{1}$ for the relaxed model restricted to $[-0.005, 0.005]$, demonstrating the recovery of the $\varphi_2$ structural layout and corner sign orientation.} 
\label{fig:compare}
\end{figure}

\subsection{Circular well}

Consider a circular well $\Omega:=\{r\in[0,1],\theta\in[0,2\pi]\}$ so that
\begin{align}
    q_{1,b} = -\tfrac{1}{2}\cos2\theta, \quad q_{2,b}=-\tfrac{1}{2}\sin2\theta \quad \text{on} \quad \partial\Omega.
\end{align}

Figure~\ref{fig:circle} shows two representative configurations. In the first row, there are two $+\tfrac{1}{2}$-defects positioned diametrically opposite along the boundary. In the second row, there is a single $+1$-defect located at the centre of the well. The former configuration has been recently observed in confined populations of spindle cells~\cite{duclos2017topological,miyazako2022explicit}. As discussed previously, the high symmetry of the circular geometry reduces the influence of the boundary conditions in the large-domain limit $\ep \gg 1$, so both mixed and Dirichlet conditions produce similar results. However, in the small-domain limit $(\ep \sim 1)$, the nematic order can melt under mixed boundary conditions, resulting in a vanishing scalar order parameter. For $\ep \ll 1$, we have
\begin{align}
 q_{1} = -\tfrac{1}{2}s(r)\cos2\theta, &\quad q_{2}=-\tfrac{1}{2}s(r)\sin2\theta ,\\
    s(r)= \frac{\ep\gamma}{2} r^2 -\frac{\ep^2\gamma^2}{4}r^2 &+ \frac{\ep^3}{24}\left[(3\gamma^3+2\gamma)r^2-\gamma r^4\right]+\cdots.
\end{align}

\subsection{Illustration of the role of surface energy in the Landau--de Gennes correction to the Oseen--Frank limit}

One of the main predictions of this study is the demonstration that including the surface anchoring energy yields an $O(1/\ep)$ director-field correction to the Oseen--Frank limit. To directly validate this prediction, we compute the full numerical solution for the diagonal state in a square well at a large domain size ($\ep = 50$) using both our relaxed surface anchoring model with $\gamma = 1$ and the rigid Dirichlet boundary condition.

We isolate the non-equilibrium structural deviations by plotting the director angle correction, defined as $\Delta \varphi = \varphi - \varphi_{\text{OF},0}$, where $\varphi = \frac{1}{2}\arctan(q_2/q_1)$ is the numerically computed director angle and $\varphi_{\text{OF},0}$ is the leading-order Oseen–Frank solution obtained from~\eqref{OFleading}--\eqref{OFleadingBC}. To reveal the underlying bulk patterns, the colour contours must be bounded within distinct limits; otherwise, the strong local singularities at the geometric corners ($\varphi_{\text{OF},k} \sim O(r^{-k})$) would completely dominate the colour scale and obscure the interior bulk fields. 

In the top-left panel of Figure~\ref{fig:compare}, which corresponds to the finite surface anchoring condition, the colour bar is restricted to the range $[-0.1, 0.1]$. In contrast, for the rigid Dirichlet case in the top-right panel, the colour bar must be restricted to a much narrower range of $[-0.001, 0.001]$ to resolve any structural variation in the bulk. The fact that the interior patterns emerge at an amplitude of $O(10^{-1})$ for the relaxed model but are restricted to $O(10^{-3})$ for the Dirichlet case provides direct quantitative confirmation that the finite surface energy drives a dominant $O(1/\ep)$ director correction, whereas the Dirichlet boundary condition suppresses this response to an $O(1/\ep^2)$ effect. These two interior patterns represent the spatial profiles of the first-order $\varphi_1$ and second-order $\varphi_2$ asymptotic corrections, respectively, with $\varphi_1$ closely matching the profile depicted in Figure 1. Furthermore, the sign of the pattern near the corners is flipped between the two models, demonstrating that they represent fundamentally different structural relaxation pathways from the Oseen--Frank solution.

To further establish the validity of the matched asymptotic framework, the bottom panel of Figure~\ref{fig:compare} isolates the second-order correction from the relaxed model by plotting $\varphi - \varphi_{\text{OF},0} - \ep^{-1}\varphi_{1}$. When the dominant $O(1/\ep)$ contribution is removed, the remaining field displays a structure that is qualitatively similar to the second-order pattern $\varphi_2$ extracted directly from the rigid Dirichlet condition, with the corner signs being restored. This figure is restricted to the narrow colour range of $[-0.005, 0.005]$.

\section{Concluding remarks}

This work has revisited the modelling of strong anchoring in confined nematic liquid crystals within the Landau--de Gennes framework. By retaining a finite surface energy and adopting the scaling $\gamma=\xi/l_{ex}\sim O(1)$, we have derived a more physical description. Key findings include
\begin{itemize}
    \item In the small-domain limit $(\ep \to 0)$, the equilibrium state is uniquely determined by the boundary-average of the anchoring data. For symmetrically frustrated geometries, this average can vanish, leading to a melted isotropic phase at leading order, with nematic order emerging only at higher orders. This contrasts with Dirichlet conditions, which enforce order even in unphysically small domains.
    \item In the large-domain (Oseen--Frank) limit $(\ep\to\infty)$, the inclusion of surface energy introduces an $O(1/\ep)$ correction to the director field near boundaries, a result not captured by Dirichlet conditions. This correction reflects a partial relaxation of the director in response to bulk elastic stresses and is most pronounced in geometrically frustrated configurations.  
    \item The structure of defect cores—both interior and boundary—is sensitively influenced by the choice of boundary condition. Mixed (Robin) conditions yield smooth, physically realistic defect profiles, whereas Dirichlet conditions produce artificially distorted cores. This is particularly evident in numerical simulations of square wells with tangential anchoring.
    \item The revised theory provides a consistent framework for studying nematic equilibria in nanoscale systems, where domain size, coherence length, and extrapolation length are comparable. It reconciles the mathematical convenience of Dirichlet conditions with the physical necessity of finite anchoring energy.
\end{itemize}

In summary, the strong anchoring condition should be understood as a distinguished limit where surface energy remains finite and competes with bulk elasticity. The proposed approach yields more accurate predictions for director configurations, defect structures, and stability in confined nematics, with implications for the design and analysis of liquid crystal-based micro- and nano-devices. Future work could extend this analysis to three-dimensional geometries, dynamic phenomena, and systems with degenerate or patterned anchoring.

\section*{Acknowledgment}
The author acknowledges helpful discussions with Apala Majumdar and Eugene C. Gartland.

\bibliographystyle{plain}
\bibliography{reference}


\appendix

\section*{Appendix A: Formulation of the stability analysis}
\label{sec:appenstability}

Let $\bar{\mb q}=(\bar q_1,\bar q_2)^T$ be a critical point of the free energy functional $F[\mb Q]$ given in~\eqref{freeenergy}. To examine its stability, we study the associated gradient–flow dynamics
\begin{equation}
    \pfr{\mb q}{t} = -\frac{\delta F}{\delta \mb q}, \quad \text{where} \quad \mb q= \begin{bmatrix}
        q_1 \\ q_2
    \end{bmatrix}
\end{equation}
and $t$ is a fictitious time that drives the system towards lower energy states. By superposing small perturbations to $\bar{\mb q}$ of the form,
\begin{equation}
   \mb q = \bar{\mb q}(x,y) + \hat{\mb q}(x,y) e^{-\sigma t},
\end{equation}
with $|\hat{\mb q}|\ll | \bar{\mb q}|$ and linearising the gradient-flow equations, we obtain the eigenvalue problem
\begin{equation}
    \mb H_{\bar{\mb q}}  \hat{\mb q}  = \sigma \hat{\mb q}  , \qquad   \mb H_{\bar{\mb q}} =  -2\mb I \nabla^2 +2\ep^2[8\bar{\mb q}\otimes\bar{\mb q} + (4|\bar{\mb q}|^2-1)\mb I ].\label{eigenvalue}
\end{equation}
The perturbations are required to satisfy the homogeneous boundary condition
\begin{equation}
    \pfr{\hat{\mb q}}{\nu} = - \ep\gamma \hat{\mb q}  \quad \text{on} \quad \partial\Omega.
\end{equation}
Here, $\mb H_{\bar{\mb q}}$ is the Hessian operator of the reduced Landau--de Gennes energy~\eqref{freeenergy} evaluated at the critical point and is defined via the second variation
\begin{equation}
\delta^2 F[\bar{\mb q}](\gb\eta,\gb\psi)
= \int_\Omega \gb\eta(\mb x)^T  \mb H_{\bar{\mb q}} \gb\psi(\mb x)  d^2x,
\end{equation}
where $\gb\eta,\gb\psi \in H^1(\Omega;\mathbb R^2)$ are admissible test functions. Equivalently, one could formally express this in terms of the functional second derivative with respect to the vector field $\mb q$ using the double-integral form
\begin{align}
    \delta^2 F[\bar{\mb q}](\gb\eta,\gb\psi) 
= \iint_\Omega 
\gb\eta(\mb x)^T 
\left. \frac{\delta^2 F}{\delta \mb q(\mb x)\delta \mb q(\mb y)} \right|_{\bar{\mb q}} 
 \gb\psi(\mb y)  d^2y  d^2x
\\ \text{with} \qquad \frac{\delta^2 F}{\delta \mb q(\mb x)\delta \mb q(\mb y)} 
= \mb H_{\bar{\mb q}}  \delta(\mb x-\mb y),
\end{align}
which represents the entire Hessian operator acting on the space of vector fields. Since $ \mb H_{\bar{\mb q}}$ is self-adjoint, all its eigenvalues are real. The spectrum is discrete and may be arranged in non-decreasing order, $\sigma_1 \leq \sigma_2 \leq \sigma_3 \leq \dots$. The critical point $\bar{\mb q}$ is asymptotically stable when $\sigma_1>0$, marginally stable when $\sigma_1=0$ and unstable when $\sigma_1<0$. The number of negative eigenvalues gives the Morse index of the equilibrium and quantifies the number of linearly unstable directions in the energy landscape.

\end{document}